\def\woodmciintyreBBI{\mathbf{B}_{\rm i}}
\def\woodmciintyreBB{\mathbf{B}}
\def\woodmciintyreBIr{B_{{\rm i}r}}
\def\woodmciintyreBIz{B_{{\rm i}z}}
\def\woodmciintyreBIphi{B_{{\rm i}\phi}}
\def\woodmciintyreBIOCROSSOVER{B_{{\rm i}0{\rm crossover}}}
\def\woodmciintyreBIO{B_{{\rm i}0}}
\def\woodmciintyreCI{C_{{\rm i}}}
\def\woodmciintyreOMEGAI{\Omega_{\rm i}}
\def\woodmciintyreN{N}
\def\woodmciintyrerp{r_{\rm p}}
\def\woodmciintyrebu{\mathbf{u}}
\def\woodmciintyreafunny{a}
\def\woodmciintyrebfunny{b}
\def\woodmciintyreUCRIT{U_{\rm crit}}
\def\woodmciintyreRossby{{\cal R}}

\def\woodmciintyreCZ{convec\-tion zone}
\def\woodmciintyreTC{ta\-cho\-cline}
\def\woodmciintyreTP{ta\-cho\-pause}
\def\woodmciintyreHSL{hel\-ium sett\-ling lay\-er}
\def\woodmciintyreHL{hel\-ium lay\-er}
\def\woodmciintyreCL{con\-fine\-ment layer}
\def\woodmciintyreCHL{con\-fine\-ment-layer}
\def\woodmciintyreCapCHL{Con\-fine\-ment-layer}
\def\woodmciintyreRH{right-hand}

\def\woodmciintyresliver{{\hskip 0.6pt}}
\def\woodmciintyreantisliver{{\hskip -0.6pt}}
\def\woodmciintyreSliver{{\hskip 1.2pt}}
\def\woodmciintyreAntisliver{{\hskip -1.2pt}}

\def\woodmciintyreAstronAstroph{A\woodmciintyresliver\&A}

\def\woodmciintyreMm{\woodmciintyreSliver Mm\woodmciintyresliver}
\def\woodmciintyrecms{\woodmciintyresliver cm\,s$^{-1}$}
\def\woodmciintyrecmms{\woodmciintyresliver cm$^2$s$^{-1}$}
\def\woodmciintyrecmmss{\woodmciintyresliver cm$^2$s$^{-2}$}
\def\woodmciintyrecmmmmss{\woodmciintyresliver cm$^4$s$^{-2}$}
\def\woodmciintyresm{{\woodmciintyresliver s$^{-1}$}}
\def\woodmciintyrepr{\woodmciintyresliver'}
\def\woodmciintyreprr{\woodmciintyresliver''}
\def\woodmciintyresubr{{\woodmciintyreantisliver r}}
\def\woodmciintyresubz{{\woodmciintyreantisliver z}}
\def\woodmciintyresubphi{{\woodmciintyreantisliver\phi}}

\documentclass[
    ,final
a4paper]
{aipproc}

\layoutstyle{6x9}

\usepackage{amsmath}
\usepackage{amssymb}

\begin{document}

\title{
   Confinement of the Sun's interior magnetic field:\\some
   exact
   boundary-layer solutions
}
  


\author{T. S. Wood}{
address={Department of Applied Mathematics and Theoretical Physics,
University of Cambridge;\\
{www.damtp.cam.ac.uk/user/tsw25/}
\ and \
{www.atm.damtp.cam.ac.uk/people/mem/}
} }

\author{M. E. McIntyre}{
address={Department of Applied Mathematics and Theoretical Physics,
University of Cambridge;\\
{www.damtp.cam.ac.uk/user/tsw25/}
\ and \
{www.atm.damtp.cam.ac.uk/people/mem/}
} }

{\footnotesize
\begin{center}
\noindent
Proc.\ July 2007 Conf. on
\emph{Unsolved Problems in Stellar Physics},
Amer.\ Inst.\ of Physics, in press.
\end{center}
}
\vspace{-1.05cm}

\begin{abstract}
High-latitude laminar
confinement of the interior field
$\woodmciintyreBBI$
is shown to be possible.
Mean downwelling $U$
as weak as
$2\times10^{-6}$\woodmciintyrecms
\! -- gyroscopically pumped by turbulent stresses in the
overlying
\woodmciintyreCZ\ and/or \woodmciintyreTC\ \,--\,
can hold the field in
advective--diffusive balance within a \woodmciintyreCL\ of
thickness scale $\delta\sim1.5$\woodmciintyreMm\
$\sim 0.002\woodmciintyresliver R_{\odot}$.
The \woodmciintyreCL\
sits at the base
of the
high-latitude \woodmciintyreTC,
near the top of the
radiative envelope and
just above the {`}\woodmciintyreTP{'} marking the
top of the \woodmciintyreHSL.
A family of exact,
laminar, frictionless, axisymmetric
\woodmciintyreCHL\ solutions
is
obtained
in cylindrical polar coordinates,
for uniform downwelling
in the limit of
strong rotation $\woodmciintyreOMEGAI$  and
stratification $\woodmciintyreN$. \
The downwelling
cannot penetrate
the \woodmciintyreHL\
and must therefore feed
into
an equatorward flow
immediately above the \woodmciintyreTP.
The retrograde
Coriolis force on that flow is balanced by a
prograde Lorentz force
within the \woodmciintyreCL.
Buoyancy forces keep the \woodmciintyreTP\ approximately horizontal.
For typical solar $\woodmciintyreN$ values $\sim10^{-3}$\woodmciintyresm\
this type~of dynamics holds over
a substantial range of colatitudes,
e.g.\ nearly out to colatitude 40$^\circ$ \woodmciintyreAntisliver when
$U \lesssim\woodmciintyresliver 10^{-5}$\woodmciintyrecms\ \woodmciintyreAntisliver for
modest $|\woodmciintyreBBI|$ values $\sim$ tens of gauss.

The angular-momentum budget
implied by the downwelling and equatorward flow,
importing low and exporting high angular momentum,
dictates that the \woodmciintyreCL\ must exert
a net retrograde torque on its surroundings through laminar Maxwell stresses.
Some of that torque is exerted downward
through the \woodmciintyreTP\ upon the interior,
against the Ferraro constraint, and
the rest is exerted
across the periphery of the
\woodmciintyreCL\
at some outer colatitude
$\lesssim 40^\circ$.
The profiles
of velocity and magnetic field
within the \woodmciintyreCL\ are fixed by two external conditions,
first the partitioning of the torque between the
contributions exerted on the interior and across
the periphery,
and second the vertical
profile of Maxwell stress
at the periphery.
In default of detailed models of
what happens
near
the  periphery,
we provisionally suggest that a natural simplest choice of model
would be one in which all the net torque is exerted
on the interior.

\end{abstract}

\maketitle

\vspace{-1.2cm}

\section{Introduction}
\vspace{-0.1cm}

 \def\rsm{\mbox{\thinspace rad\vspace{2pt} s$^{-1}$}}
The near-rigid rotation
$\woodmciintyreOMEGAI=2.7\times10^{-6}$\rsm\
observed in the Sun's interior can be
most credibly explained via
the Ferraro constraint from a confined
global-scale interior magnetic field $\woodmciintyreBBI$
\citetext{Gough \& McIntyre 1998, hereafter GM98};
also McIntyre~\citetext{1994--2007}.     
For stability $\woodmciintyreBBI$        
must have
comparable toroidal and poloidal components
\citetext{e.g.\ Braithwaite \& Spruit 2004}. \
$\woodmciintyreBBI$~could be
axisymmetric and aligned with the Sun's rotation axis as proposed in
GM98, or oblique as is typical of Ap~stars.
We focus on the aligned case as presenting, in some ways,
the greatest problems.

The main problem, previously addressed
by Garaud~\citetext{2002--\woodmciintyresliver 08}
and by \cite{BZ06},
is how to confine $\woodmciintyreBBI$ at the pole and
in high latitudes.
It is necessary
to stop the poloidal field from diffusing up through the
polar caps and thereby
imposing the \woodmciintyreCZ's
high-latitude differential rotation upon the interior.
Such differential rotation
conflicts with the helioseismic evidence.

\begin{figure}
\resizebox{0.91\textwidth}{!}
{\includegraphics [clip=true,bb = -130 360 766 530, height=8.0cm]
{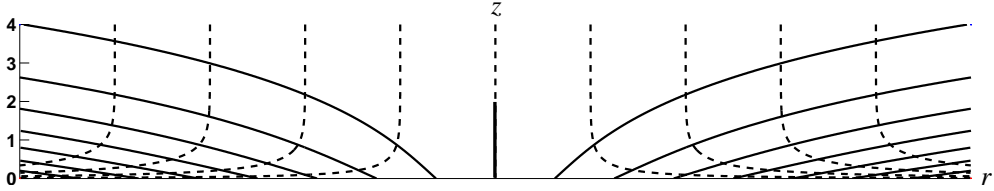}}
\caption{\footnotesize
Streamlines of the downwelling flow \,$\woodmciintyrebu$\, (dashed curves) and
of the poloidal part of \,$\woodmciintyreBB$\, (solid curves), spacing
arbitrary.
The horizontal axis is colatitude in arbitrary units, measured from
the central symmetry axis or rotation axis.  The
vertical axis is heliopotential altitude, or radial increment, $z$, \
in units of $\delta$. \
Although there is
a field line on the central symmetry axis, as well as a streamline,
the strength of the field decays
like $\exp(-z/\delta)$
with altitude $z$.
The downwelling profile $u_\woodmciintyresubz (z)=-U[1-\exp(-3z/\delta)]$.
\vspace{-0.5cm}
}
\label{fig:stream}
\end{figure}

Following
GM98 we propose that $\woodmciintyreBBI$ is confined
to the interior,
in high latitudes,
via a frictionless
laminar boundary layer at the
base
of the stably stratified tachocline.
The confinement is brought about
by
weak
downwelling of magnitude $U$, say,
taken to be
a persistent feature of the
mean meridional circulations (MMCs)
in the lower \woodmciintyreTC.
Such downwelling is to be expected
from the `gyroscopic pumping'
caused by turbulent
Reynolds and Maxwell stresses in the overlying
layers
\citep{M07},
in the same way that Ekman pumping is caused
by ordinary viscous stresses.

The overlying turbulent layers
consist of the \woodmciintyreCZ\
and possibly also the \woodmciintyreTC\
\citep{S02}.
There are uncertainties in how to
characterize those turbulent layers
in detail.
But because of the
{`}antifrictional{'}
sense
of the \woodmciintyreCZ's
turbulent
stresses --  driving
it retrogradely relative to the interior
in high latitudes --
there should be a
systematic
tendency, by one means or another,
for
the
gyroscopic pumping to produce downwelling
above the top of the radiative envelope
in high latitudes.
One possible
such scenario is discussed
in
\cite{M07},
following
\cite{S02}.
There the
gyroscopic pumping is,
in fact,
a case of ordinary Ekman pumping
near the bottom of an MHD-turbulent \woodmciintyreTC.

We assume that the bulk of the
radiative envelope
beneath
is itself
locked into rigid rotation $\woodmciintyreOMEGAI$
with the help of
the Ferraro
constraint from $\woodmciintyreBBI$,
and that gravitational settling has
produced molecular-weight gradients in the form of
a \woodmciintyreHSL\ in the outer 50\woodmciintyresliver--\woodmciintyreantisliver100\woodmciintyreMm\ or so.
Such a helium layer
is a feature of standard solar models
\citep{CDIR97}.
Its existence is indicated also by the
helioseismic evidence, despite
current uncertainties
about heavy-element abundances and their effects on opacity
\citep{CDM07}.

The
helium diffusivity is tiny, $\sim 10$\woodmciintyrecmms,
in comparison with
the thermal and magnetic diffusivities
$\kappa\sim10^7$\woodmciintyrecmms\ and
$\eta\sim3\times10^2$\woodmciintyrecmms\ 
\citetext{e.g.\ Gough 2007}.
Therefore the \woodmciintyreHL\
is
nearly
impervious to
MMCs.
\cite{MeM86}
call this the {`}$\mu$-choke{'} or  {`}$\mu$-barrier{'} effect;
see also
\cite{Me53}.
The high-latitude
downwelling, whatever its origin,
must therefore
feed
into
an equatorward flow just above
the {`}\woodmciintyreTP{'}
marking the top of the \woodmciintyreHL.
The retrograde Coriolis force~on that equatorward flow
needs to be
balanced by a prograde Lorentz force.
Buoyancy forces
from the stable stratification
keep the \woodmciintyreTP\ and the \woodmciintyreHL\ beneath it very
close to the horizontal,
along with the  stratification surfaces themselves
\citetext{McIntyre 2007, \S8.5}.
We report
a new family of
exact
steady
solutions of the nonlinear equations
showing how all these elements
fit together,
confining $\woodmciintyreBBI$
with\-in
a layer of thickness scale
$\;\delta=\eta/U\,$
while
transmitting a
retrograde torque to
the interior.

\vspace{-0.29cm}

\section
        {\woodmciintyreCapCHL\ SOLUTIONS}

\vspace{-0.25cm}

The \woodmciintyreCHL\ solutions are
axisymmetric similarity solutions obtained
in cylindrical polar coordinates $(r,\,\phi,\,z)$ and
valid in some region surrounding the pole,
for any finite stratification $\woodmciintyreN$. \
Figure~\ref{fig:stream} shows an example, with the
north
pole at the centre.
The $z$~axis
is central,
pointing upward,
and
$z$ measures altitude in units of $\delta$. \
The cylindrical
radial
coordinate
$r$
is proportional to colatitude and
is in arbitrary units,
the
structure
being
self-similar under radial dilatation.
The similarity
solutions
have vertical field components
$u_\woodmciintyresubz $ and $B_\woodmciintyresubz $ that
are independent of $r$, and horizontal components
$u_\woodmciintyresubr $, $u_\woodmciintyresubphi $ and $B_\woodmciintyresubr $, $B_\woodmciintyresubphi $
proportional to $r$. \
Because $\nabla\cdot\woodmciintyreBB=0$
we have $2B_\woodmciintyresubr  = -B\woodmciintyrepr_\woodmciintyresubz  r$
where the prime denotes
$\partial\!/\woodmciintyreAntisliver\partial z$. \
We assume anelastic flow
with
$\delta \ll
$ pressure scale height (60\woodmciintyreMm\ or more),
so that background density is
constant (Boussinesq limit) and
$\nabla\cdot\woodmciintyrebu=0$, implying
$2u_\woodmciintyresubr  = -u\woodmciintyrepr_\woodmciintyresubz r$.

\begin{figure}
{\includegraphics [width = 7.5cm, clip=true, bb = 125 62 910 652]
{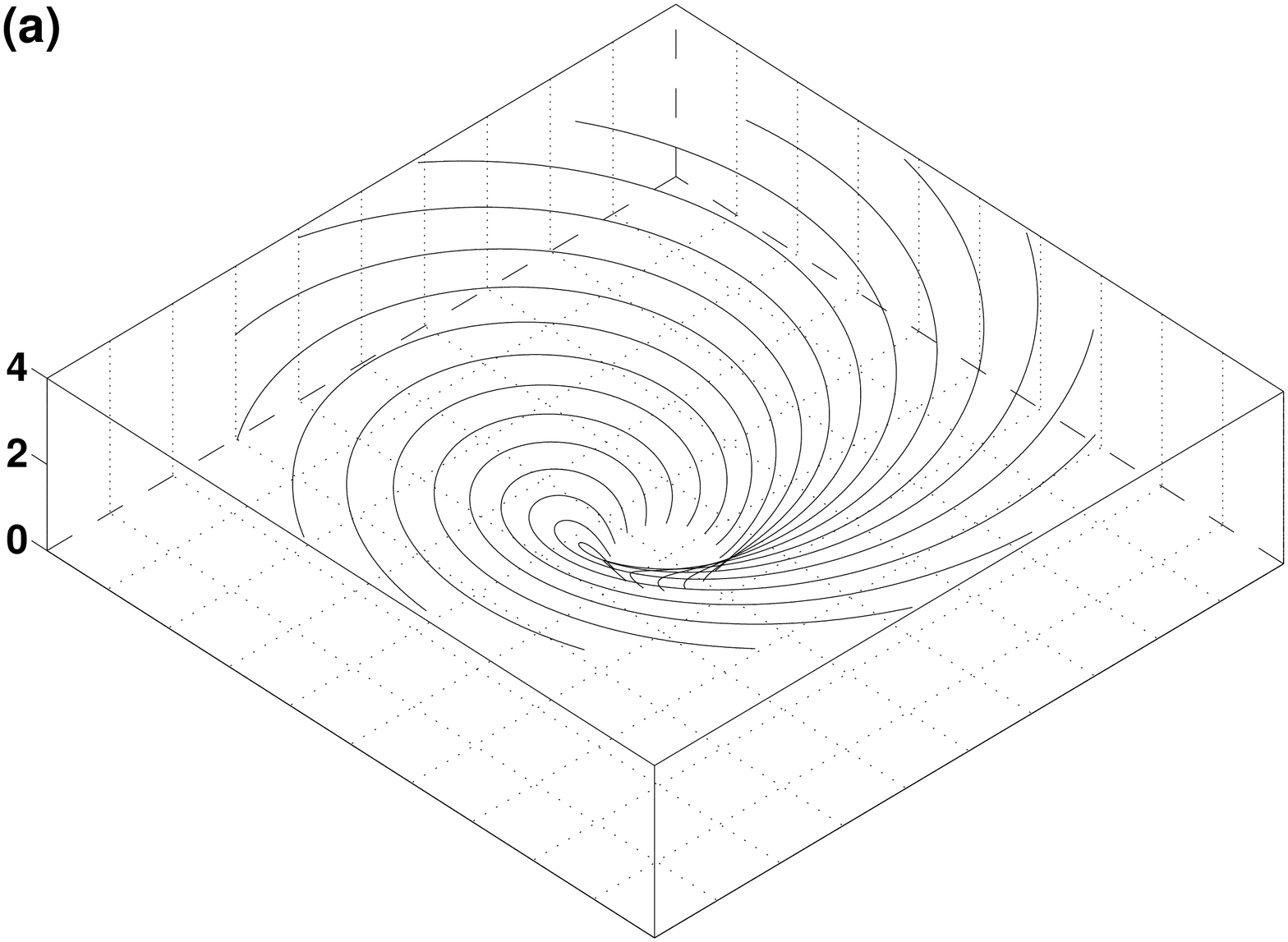}}
\hspace{0.5cm}
{\includegraphics [clip=true, bb = -41 80 622 721,
width=6.0cm]
{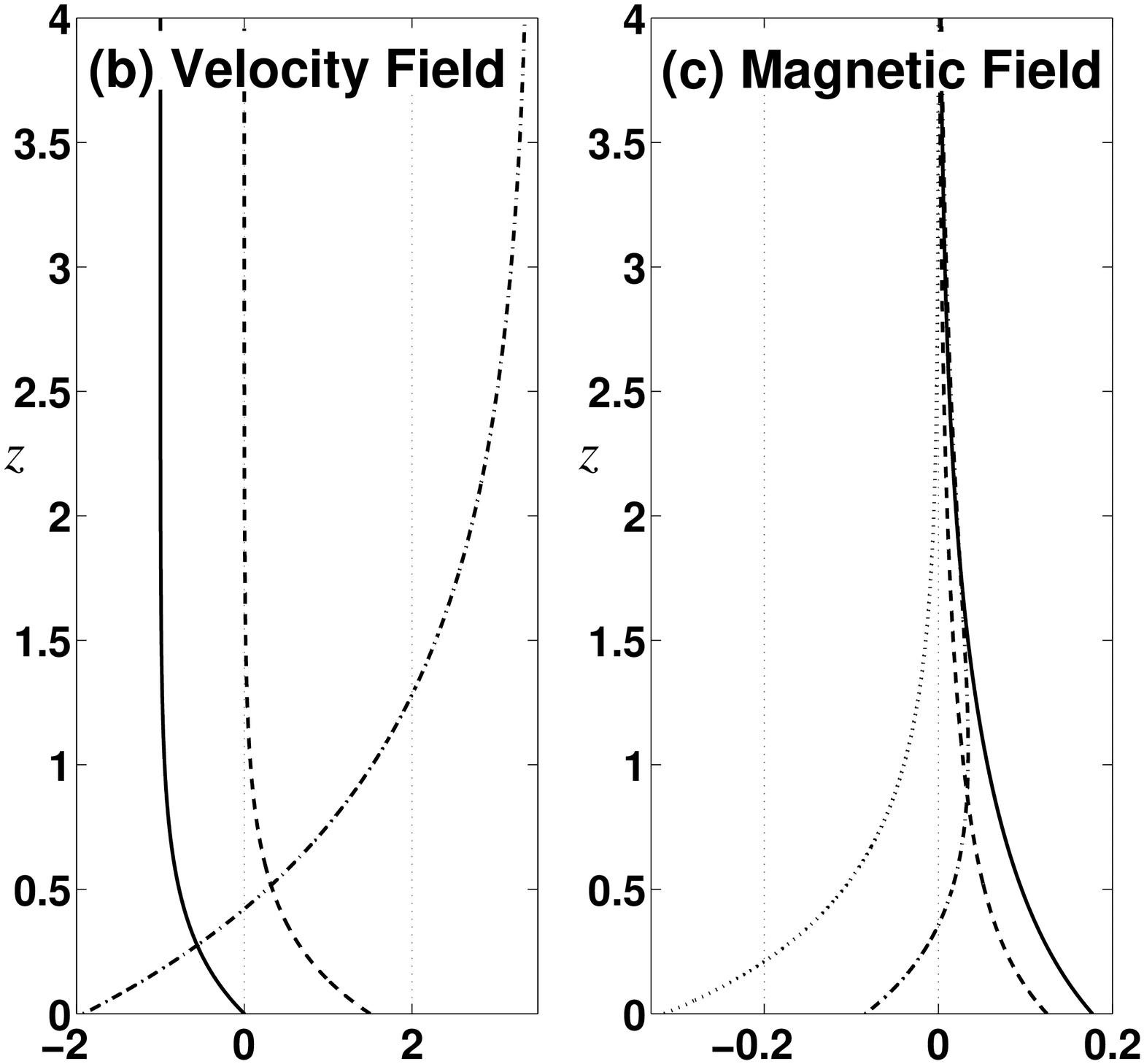}}
\caption{\footnotesize
Further views of the same solution.  The
view (a)
of field lines
shows only those lines that emerge from
the \woodmciintyreTP\ at a single colatitude near the pole.  The lines emerging
at other colatitudes have the same shape dilated horizontally
(see text).
The profiles (b) are those of
$u_\woodmciintyresubz $ (solid), $u_\woodmciintyresubr $ (dashed) and 
$u_\woodmciintyresubphi$
(dot-dashed), in units of
$U$\!,\,
$U\woodmciintyreantisliver r/\delta$ and
$\woodmciintyreSliver 50\woodmciintyresliver U\woodmciintyreAntisliver r/\delta$ respectively;
(c) are
$B_\woodmciintyresubz$ (solid), $B_\woodmciintyresubr $ (dashed) $B_\woodmciintyresubphi$ (dot-dashed), and
the aux\-iliary function $\hat B_\woodmciintyresubphi$ (dotted), in units of
$(\eta\woodmciintyreOMEGAI)^{1/2}$ for $B_z$ and
$(\eta\woodmciintyreOMEGAI)^{1/2}r/\woodmciintyreantisliver\delta$ for the rest.
The shape of
$\hat u_\woodmciintyresubphi(z)$,
not shown in (b), is
nearly
indistinguishable from
$u_\woodmciintyresubphi(z) - u_\woodmciintyresubphi(\infty)$. \
See (\ref{eq:hatBphi})\woodmciintyresliver --\woodmciintyresliver(\ref{eq:beta})
below.
\vspace{-0.5cm}
}
\label{fig:profiles}
\end{figure}

The boundary conditions
include continuity of $\woodmciintyreBB$ at the \woodmciintyreTP\ $z=0$.
The magnetic field $\woodmciintyreBBI$ just beneath the \woodmciintyreTP\
therefore has
a matching similarity structure.
The simplest such field has
$\woodmciintyreBIr = \woodmciintyreBIO r$ and $\woodmciintyreBIz = -2\woodmciintyreBIO\woodmciintyresliver z
+ \woodmciintyreCI$, where $\woodmciintyreBIO$ and $\woodmciintyreCI$ are constants, so that $\nabla\cdot\woodmciintyreBBI = 0$,
while the azimuthal component
$\woodmciintyreBIphi$ is taken
$\propto\woodmciintyreBIz r$
to ensure the vanishing of the interior's
azimuthal Lorentz force, \woodmciintyreRH\ side of (\ref{eq:mom-phi}) hereunder.
Figure \ref{fig:profiles} shows further views of the same solution.
The equations solved are
\vspace{-0.02cm}
\begin{eqnarray}
  \eta B\woodmciintyreprr_\woodmciintyresubz 
&=& 
  B\woodmciintyrepr_\woodmciintyresubz  u_\woodmciintyresubz  - B_\woodmciintyresubz u\woodmciintyrepr_\woodmciintyresubz
\label{eq:ind-z}
\\
  \eta B\woodmciintyreprr_\woodmciintyresubphi 
&=& 
  B\woodmciintyrepr_\woodmciintyresubphi u_\woodmciintyresubz  - B_\woodmciintyresubz u\woodmciintyrepr_\woodmciintyresubphi  
  \label{eq:ind-phi}
\\
  2\woodmciintyreOMEGAI\woodmciintyresliver u_\woodmciintyresubr
  \;=\; 
  -\woodmciintyreOMEGAI\woodmciintyresliver u\woodmciintyrepr_\woodmciintyresubz\woodmciintyresliver r
&=& 
  B_\woodmciintyresubz B\woodmciintyrepr_\woodmciintyresubphi  - B\woodmciintyrepr_\woodmciintyresubz B_\woodmciintyresubphi
  \;=\;
  B_\woodmciintyresubz ^2(B_\woodmciintyresubphi /B_\woodmciintyresubz )'
\label{eq:mom-phi}
\vspace{-0.42cm}
\end{eqnarray}
of which the first two
come from the induction equation
after substituting the similarity structure into
$\nabla\times(\woodmciintyrebu\times\woodmciintyreBB)$,
and the third from the azimuthal
momentum equation
in the limit of
small Rossby number, thus retaining
only Coriolis and not relative accelerations;
$\woodmciintyreBB$ is in units of Alfv\'en speed,
$\sim{\textstyle{1\over2}}$\woodmciintyrecms\ per gauss at \woodmciintyreTC\ mass densities.
The $\eta$
terms are exact because
the similarity structure
makes the horizontal derivatives vanish.

In the absence of specific information about
the vertical profile
of Maxwell stresses $B_\woodmciintyresubr B_\woodmciintyresubphi $ at the periphery,
there is an undetermined function of $z$ in the problem.
We may use this freedom to specify
the vertical profile $u_\woodmciintyresubz (z)$ of the downwelling.
Then (\ref{eq:ind-z}) becomes a
linear problem for $B_\woodmciintyresubz (z)$. \ It is
solvable with
$B\woodmciintyrepr_\woodmciintyresubz (0)=-2\woodmciintyreBIO$  (matching the $r$ components)
and with \woodmciintyresliver$B_\woodmciintyresubz (z)$\woodmciintyresliver\
decaying like \,${\exp}({-z/\delta})$\, as $z\to\infty\woodmciintyreSliver$. \
This determines both $B_\woodmciintyresubz (z)$ and $\woodmciintyreCI$\woodmciintyresliver.
Then,
provided only that $|u_\woodmciintyresubz |$ converges to $\woodmciintyresliver U$ faster than
${\exp}({-2z/\delta})$, \
(\ref{eq:mom-phi}) has a
solution
\vspace{-0.15cm}
\begin{equation}
~~B_\woodmciintyresubphi\:~=~~\hat B_\woodmciintyresubphi
\:~\equiv\;
-2\woodmciintyreOMEGAI\woodmciintyreSliver B_\woodmciintyresubz
\int_z^\infty
   \frac{u_\woodmciintyresubr }{B_\woodmciintyresubz ^2}\,
\woodmciintyresliver dz
~.
\label{eq:hatBphi}
\vspace{-0.08cm}
\end{equation}
In the example shown in the figures we took
$\;u_\woodmciintyresubz
\;=\;
-U[1-\exp(-3z/\delta)]
$, \
which implies that
$\;u_\woodmciintyresubr
\woodmciintyresliver=\woodmciintyreantisliver
-{\textstyle{1\over2}} u\woodmciintyrepr_\woodmciintyresubz\woodmciintyresliver r
\woodmciintyresliver=\woodmciintyresliver
{\textstyle{3\over2}} (Ur/\delta) \exp(-3z/\delta)
$,\,
giving the
$\hat B_\woodmciintyresubphi $ profile
shown as the dotted, left\-most~curve
in
Fig.~\ref{fig:profiles}c. \
Then (\ref{eq:ind-phi}) has
a corresponding solution
\vspace{-0.15cm}
\begin{equation}
~~u_\woodmciintyresubphi\:~=~~\hat u_\woodmciintyresubphi
\:~\equiv~
\int_z^\infty
   \frac{\eta\hat B\woodmciintyreprr_\woodmciintyresubphi  - u_\woodmciintyresubz \hat B\woodmciintyrepr_\woodmciintyresubphi }{B_\woodmciintyresubz }
\woodmciintyreSliver dz
~.
\label{eq:hatuphi}
\vspace{-0.13cm}
\end{equation}
Under our assumptions, both (\ref{eq:hatBphi}) and (\ref{eq:hatuphi})
are finite at $z=0\woodmciintyresliver$, and are evanescent
as $z\to\infty\woodmciintyreSliver$,\, respectively
like ${\exp}({-2z/\delta})$
and  ${\exp}({-z/\delta})$
in the example shown.  The foregoing procedure is
robust and well-conditioned.
In the simplest cases in which
$u_\woodmciintyresubz $ and $u\woodmciintyrepr_\woodmciintyresubz$ are both negative for all $z$\woodmciintyresliver,
as in the figures,\,
it is easy to see by inspection of (\ref{eq:ind-z})\woodmciintyresliver,
qualitative\-ly integrating it downward
from $z=\infty$\woodmciintyresliver,
that $B\woodmciintyreprr_\woodmciintyresubz$ on the left and both terms on the right
are positive for all $z$.  It then follows
from (\ref{eq:hatBphi}) that
$\hat B_\woodmciintyresubphi $ is negative for all $z$. \
But $\hat u_\woodmciintyresubphi $ and $\hat u\woodmciintyrepr_\woodmciintyresubphi $ can change sign,
though in fact $\hat{u}_\woodmciintyresubphi $ is negative and monotonic
in the example shown.

We still have a pair of undetermined parameters at our disposal because
(\ref{eq:ind-phi}) and (\ref{eq:mom-phi}) are
also satisfied, for any constants $\alpha$ and $\beta$, by
\vspace{-0.1cm}
\begin{equation}
B_\woodmciintyresubphi 
~=~
\hat B_\woodmciintyresubphi  + \alpha B_\woodmciintyresubz  r
\qquad\mbox{~and}\qquad
u_\woodmciintyresubphi 
~=~
\hat u_\woodmciintyresubphi  +  (\alpha u_\woodmciintyresubz  + \beta)r
~.
\label{eq:Bphi-uphi-gen}
\vspace{-0.1cm}
\end{equation}
Here (\ref{eq:ind-z})
has been used to
simplify the last term on the right.
The
$\alpha$ term in $B_\woodmciintyresubphi $
contributes nothing
to the azimuthal
Lorentz force
on the \woodmciintyreRH\ side of  (\ref{eq:mom-phi}),
but does change the
Maxwell stresses acting across the \woodmciintyreTP\ $z=0$ and the periphery,
$r=\woodmciintyrerp$ say,
by equal and opposite amounts.
In other words
$\alpha$ governs
the partitioning of Maxwell torques
   between     \woodmciintyreTP\ and     periphery.

In the example
shown $\alpha$ was chosen,
purely on Occam's-razor grounds,
to make
the
Maxwell torque
on the
periphery
zero.  Lacking information about
conditions at the periphery, zero is arguably the simplest choice.\,
The torque $\propto \int_0^\infty B_\woodmciintyresubr B_\woodmciintyresubphi \,dz$.\,
It is zero if
\vspace{-0.15cm}
\begin{equation}
\alpha
~=~
-\,
\frac{\int_0^\infty B_\woodmciintyresubr  \hat{B}_\woodmciintyresubphi \,dz}
{\int_0^\infty B_\woodmciintyresubr B_\woodmciintyresubz r\,dz}
~=~
1.31\woodmciintyresliver\delta^{-1}
\label{eq:alphadef}
\vspace{-0.15cm}
\end{equation}
from the first of (\ref{eq:Bphi-uphi-gen});\,
the quotient is independent of $r$\, because of the
similarity structure.

To
find $\beta$ we integrate (\ref{eq:ind-phi}) across the \woodmciintyreTP\
and use (\ref{eq:mom-phi}) to give
\vspace{-0.2cm}
\begin{equation}
  u_\woodmciintyresubphi (0)
~=\;
-\,
\Lambda^{-1}
u_\woodmciintyresubr (0)
\label{eq:slip-velocities}
\vspace{-0.2cm}
\end{equation}
where $\Lambda \equiv
B_\woodmciintyresubz ^2(0)/(2\woodmciintyreOMEGAI\eta)$,
the Elsasser number based on $B_\woodmciintyresubz (0)$  $(\sim \woodmciintyreBIO\delta)$,
determining
the direction of the
frictionless slip flow at the \woodmciintyreTP\
just above the
rigidly-rotating interior. \ The slip flow is
equatorward and retrograde, following
a logarithmic spiral. \
For finite viscosity
$\nu$
there would be a laminar Ekman
layer of thickness scale
$\delta_\nu\sim(\nu/\Omega)^{1/2}
\sim
3\times10^{-5}
$\woodmciintyreMm\
$\ll\delta$, \
if we take
$\nu\sim30$\woodmciintyrecmms\ 
\citep{Go07}.
Its
flow is unobstructed by the field lines since
magnetic diffusion
on the scale $\delta_\nu$ is almost instantaneous.  \
Note (\ref{eq:slip-velocities}) gives the spiral
just \emph{above} the Ekman layer.
From (\ref{eq:Bphi-uphi-gen}) and (\ref{eq:slip-velocities}),
\vspace{-0.09cm}
\begin{equation}
  \beta
~=~
- r^{-1}\hat{u}_\woodmciintyresubphi (0)
\;-\;
\Lambda^{-1}r^{-1}u_\woodmciintyresubr (0)
\:~=~\:
1.73 \woodmciintyreSliver U\delta^{-1}\! \times10^2
\label{eq:beta}
~.
\vspace{-0.2cm}
\end{equation}
In fact $\Lambda^{-1}$
measures
the spiralling
of the field lines
as well as that of the flow lines,
because
(\ref{eq:hatBphi})--(\ref{eq:beta}) imply the
order-of-magnitude relations
\vspace{-0.15cm}
\begin{equation}
  B_\woodmciintyresubphi
  ~\sim~
  \hat{B}_\woodmciintyresubphi
  ~\sim~
  2\woodmciintyreOMEGAI\woodmciintyreSliver U\woodmciintyreAntisliver r
/
  B_\woodmciintyresubz 
~,\qquad \mbox{} \qquad
  u_\woodmciintyresubphi
  ~\sim~
  \hat{u}_\woodmciintyresubphi
  ~\sim~
  U\woodmciintyreantisliver B_\woodmciintyresubphi
/
  B_\woodmciintyresubz 
~,
\label{eq:Bu-scales}
\vspace{-0.2cm}
\end{equation}
and
\vspace{-0.4cm}
\begin{equation}
  B_\woodmciintyresubphi /B_\woodmciintyresubr 
  ~\sim~
  u_\woodmciintyresubphi /u_\woodmciintyresubr 
  ~\sim~
  \Lambda^{-1}
~.
\label{eq:Bu-spirals}
\vspace{-0.05cm}
\end{equation}
Recall that $\delta=\eta/U$ and \
$2u_\woodmciintyresubr
=
-u\woodmciintyrepr_\woodmciintyresubz r \sim Ur/\delta$, \
$2B_\woodmciintyresubr
=
-B\woodmciintyrepr_\woodmciintyresubz r \sim \woodmciintyreBIO r$. \
The numerical factors implicit in (\ref{eq:Bu-spirals}) differ
considerably from unity because of the disparity in vertical scales
between $\exp(-z/\delta)$, $\exp(-2z/\delta)$ and
$\exp(-3z/\delta)$,
along with
the peculiar balance of terms in (\ref{eq:ind-phi})
that enables $\hat{B}_\woodmciintyresubphi $ to
evanesce faster than  $\exp(-z/\delta)$.
In the example shown in the figures,
$\Lambda =
1.57\times10^{-2}\!$. \ \
Fig.~\ref{fig:profiles} shows that
$|B_\woodmciintyresubphi /B_\woodmciintyresubr |$ attains values considerably smaller numerically,
and $|u_\woodmciintyresubphi /u_\woodmciintyresubr|$ distinctly larger,
than $\Lambda^{-1}\approx60$.

By contrast with GM98's thermomagnetic boundary layer,
whose dynamics crucially involved the tilting of stratification surfaces,
our exact solutions of (\ref{eq:ind-z})--(\ref{eq:mom-phi})
impose no restriction on $U$ values and
mass throughput for given $\woodmciintyreBIO$. \
However, there is an implicit restriction,
for given peripheral radius $r=\woodmciintyrerp$
and
stratification $\woodmciintyreN$. \
$N$ has been
assumed strong enough to hold stratification surfaces horizontal.
Only then can the uniform downwelling satisfy
the thermal diffusion equation,
displacing the stratification surfaces vertically
without tilting them.
A scale analysis, omitted for brevity, shows that
the tilting
becomes
noticeable at the periphery if
in order of magnitude
\vspace{-0.2cm}
\begin{equation}
  \quad   U
  ~\sim~
  \woodmciintyreUCRIT
  ~\equiv~
  \min
  \left[
     (\woodmciintyreafunny\woodmciintyreSliver\woodmciintyreBIO/\woodmciintyrerp)^{1/3}
     ,~~
     \woodmciintyrebfunny\woodmciintyreSliver(\woodmciintyreBIO\woodmciintyrerp)^{-1}
  \right]
\vspace{-0.25cm}
\label{eq:min-reln}
\end{equation}
where
$
  \woodmciintyreafunny
  =
  0.4
  (\eta/\!\kappa)^{1/2}
  (\eta^2\woodmciintyreN/\woodmciintyreOMEGAI)
  \sim 0.7\times10^5
$\woodmciintyrecmmmmss\ and
$
 \woodmciintyrebfunny
 =0.1
 (\eta/\!\kappa)^{1/2}\eta\woodmciintyreN
 \sim 0.15\times10^{-3} 
$\woodmciintyrecmmss.
The
min function
arises from the azimuthal vorticity balance.  In the strong-field
case (second argument, roughly corresponding to 
$\Lambda\gtrsim1$), the tilting of
the stratification surfaces is balanced solely by a Lorentz
force-curl.
In the weak-field case (first argument, $\Lambda\lesssim 1$) there is
an additional contribution from vortex twisting 
$2\woodmciintyreOMEGAI u\woodmciintyrepr_\woodmciintyresubphi$. \
The crossover corresponds to $\woodmciintyreBIO\woodmciintyrerp \sim$ 15 \woodmciintyrecms\
($\sim30$G) when $\woodmciintyrerp=350$\woodmciintyreMm,\break

\vspace{-0.44cm} 

\noindent
i.e.\ to $|\woodmciintyreBBI|$ $\sim30$G near an outer colatitude
$\sim40^\circ$. \ \   Then $\woodmciintyreUCRIT \sim 10^{-5}$\woodmciintyrecms.

\vspace{0.15cm}                   

The Rossby number
$\woodmciintyreRossby = \max\!
   |r^{-1}\woodmciintyreAntisliver\woodmciintyrebu\!\cdot\!\nabla
   (\woodmciintyreantisliver ru_\phi\!)/(\woodmciintyreAntisliver2\woodmciintyreOMEGAI u_r\woodmciintyreAntisliver)
   |
\!\sim\! (U\!\woodmciintyreAntisliver/\woodmciintyreAntisliver B_z(0))^2
\!\sim\! U^4\!\woodmciintyreAntisliver/\woodmciintyreAntisliver(\eta\woodmciintyreBIO)^2
$,
$\lesssim   10^{-6}\!$
at crossover if $U\!\lesssim\woodmciintyreUCRIT$,
and similarly small throughout the parameter range of interest,
strongly justifying
our use of the small-$\woodmciintyreRossby$ limit. \
Inverse \mbox{gradient} Rich\-ard\-son numbers
$|\woodmciintyrebu\woodmciintyrepr|^2/\woodmciintyreN^2$
$\sim
(U/\woodmciintyreUCRIT)
(\eta/\!\kappa)
\woodmciintyreRossby
$
in the \woodmciintyreCL,  and
$\sim
(U/\woodmciintyreUCRIT)
(\eta/\!\kappa)
\woodmciintyreRossby
(\delta/\delta_\nu)^2
\sim
(U/\woodmciintyreUCRIT)
(\eta/\!\kappa)
U^2 \woodmciintyreOMEGAI
/
(\nu\woodmciintyreBIO^2)
$
in the Ekman slip layer,
respectively
$\lesssim10^{-11}$ and $\lesssim10^{-3}$ at crossover. \
So the flows are strongly shear-stable.

\cite{S99}
shows that in stably stratified shear flows
the first MHD instabilities to kick in will be
diffusion-mediated Tayler kink or tipping
instabilities of $B_\phi$. \
In the weak-field case $\Lambda\lesssim 1$ (the most vulnerable,
with strong spiralling)
we find
Tayler
stability
for
$U\lesssim\,(\woodmciintyreBIO/\woodmciintyreBIOCROSSOVER)^{1/6}\,\woodmciintyreUCRIT\woodmciintyreSliver$. \
Stability increases further
when $\Lambda\gtrsim1$. \
So\break

\newpage
\noindent
 the solutions
probably
represent real laminar flows.

\vspace{-0.37cm}

\section{Concluding remarks}
\label{sec:conclu}

\vspace{-0.15cm}

The suggestion in (\ref{eq:min-reln})
of a limiting mass flow and therefore, by
implication, of an upper bound on
the torque transmissible to the interior,
is no more than a suggestion at
present. \
However, the
scaling leading to (\ref{eq:min-reln}) does
have points of similarity
to
the scal\-ing governing the mass-flow-limited thermomagnetic
boundary layer proposed in GM98.
A peripheral thermomagnetic boundary layer
might impose a mass-flow limit.
Such a limit would
have
implications, in turn, for
the possible range of interior field strengths $|\woodmciintyreBBI|$. \
In particular, the
steep falloff of $\woodmciintyreUCRIT$ on the strong-field side
of (\ref{eq:min-reln}) suggests a sharp upper bound on
confinable $|\woodmciintyreBBI|$ strengths.

A mass-flow limit,
if confirmed, would also bear on
the question of whether a
Tayler--Spruit dynamo can
run continuously or intermittently in the \woodmciintyreTC\ above the \woodmciintyreCL\
\citetext{McIntyre 2007, \S8.4}.
That question is critical to associated questions about
deep \woodmciintyreTC\ ventilation and lithium burning.

\vspace{-0.29cm}

\begin{theacknowledgments}
We thank
J\o rgen Christensen-Dalsgaard, Werner D\"appen,
Scilla Degl'Innocenti, Pascale Garaud, Douglas Gough,
Mark Miesch,
Steven Shore, and Mike Thompson for helpful comments. \
TSW is supported by a Research
Studentship from the Science and Tech\-nology Facilities Council.

\vspace{-0.85cm}

\end{theacknowledgments}


\begin{thebibliography}{dummy-label}




\bibitem[Braithwaite \& Spruit(2004)]{BS04}
Braithwaite J., Spruit H. C., 2004,
Nat, 431, 819

\bibitem[Brun \& Zahn(2006)]{BZ06}
Brun A. S., Zahn J.-P., 2006,
\woodmciintyreAstronAstroph, 457, 665

\bibitem[e.g.\ Ciacio et~al.(1997)]{CDIR97}
Ciacio F., Degl'Innocenti S., Ricci B., 1997:
\woodmciintyreAstronAstroph, 123, 449

\bibitem[e.g.\ Christensen-Dalsgaard \& Di Mauro(2007)]{CDM07}
Christensen-Dalsgaard J., Di Mauro M. P., 2007,
in Straka C. W., Lebreton Y., Monteiro  M. J. P. F. G., eds,
Stellar Evolution and Seismic Tools for Asteroseismology --
Diffusive Processes in Stars and Seismic Analysis.
EAS Publ.\ Ser.\ 26, EDP Sciences, Les Ulis, France,
DOI: 10.1051/eas:2007121

\bibitem[Garaud(2002)]{Ga02}
Garaud P., 2002,
MNRAS, 329, 1

\bibitem[Garaud(2003)]{Ga03}
Garaud P., 2003,
in Thompson M. J., Christensen-Dalsgaard, J., eds,
Stellar Astrophysical Fluid Dynamics.
Cambridge University Press, Cambridge

\bibitem[Garaud(2007)]{Ga07}
Garaud P., 2007,
in Hughes D. W., Rosner R., Weiss N. O., eds,
The Solar Tachocline.
Cambridge University Press, Cambridge

\bibitem[Garaud(2008)]{Ga08}
Garaud P., 2008, this Proceedings.

\bibitem[Gough(2007)]{Go07}
Gough D. O., 2007,
in Hughes D. W., Rosner R., Weiss N. O., eds,
The Solar Tachocline.
Cambridge University Press, Cambridge

\bibitem[Gough \& McIntyre(1998)]{GM98}
Gough D. O., McIntyre M. E., 1998,
Nat, 394, 755

\bibitem[McIntyre(1994)]{M94}
McIntyre M. E., 1994,
in E. Nesme-Ribes, ed,
The Solar Engine and its Influence on the Terrestrial 
Atmosphere and Climate (Vol.~25 of NATO ASI Subseries I, Global 
Environmental Change),
Springer-Verlag, Heidelberg

\bibitem[McIntyre(2003)]{M03}
McIntyre M. E., 2003,
in Thompson M. J., Christensen-Dalsgaard J., eds,
Stellar Astrophysical Fluid Dynamics.
Cambridge University Press, Cambridge

\bibitem[McIntyre(2007)]{M07}
McIntyre M. E., 2007,
in Hughes D. W., Rosner R., Weiss N. O., eds,
The Solar Tachocline.
Cambridge University Press, Cambridge

\bibitem[Mestel(1953)]{Me53}
Mestel L., 1953,
MNRAS, 113, 716

\bibitem[Mestel \& Moss(1986)]{MeM86}
Mestel L., Moss D. L., 1986,
MNRAS, 221, 25


\bibitem[Spruit(1999)]{S99}
Spruit H. C., 1999,
\woodmciintyreAstronAstroph, 349, 189

\bibitem[Spruit(2002)]{S02}
Spruit H. C., 2002,
\woodmciintyreAstronAstroph, 381, 923

\end{thebibliography}
\end{document}